\documentclass[traditabstract, bibyear]{aa} 
\usepackage{graphicx}
\usepackage{txfonts}
\usepackage{color}
\usepackage{amstext}

\begin{document}

   \title{Periodic transit and variability search with simultaneous systematics filtering: Is it worth it?}

   \author{G\'eza Kov\'acs\inst{1},
           Joel D. Hartman\inst{2}, 
           G\'asp\'ar \'A. Bakos\inst{2}\fnmsep\thanks{Alfred P.~Sloan Research Fellow}\fnmsep\thanks{Packard Fellow}
          }

   \institute{Konkoly Observatory, Budapest, 1121 Konkoly Thege ut. 13-15, Hungary \\
              \email{kovacs@konkoly.hu}
	      \and
	      Department of Astrophysical Sciences, Princeton University, Princeton, NJ 08544, USA \\
             }

   \date{Received August 5, 2015; Accepted October 16, 2015}

 
  \abstract
{By using subsets of the HATNet and K2 (Kepler two-wheel) 
Campaign~1 databases, we examine the effectiveness of filtering out 
systematics from photometric time series while simultaneously searching 
for periodic signals. We carry out tests to recover simulated sinusoidal  
and transit signals added to time series with both real and artificial 
noise. We find that the simple (and more traditional) method that 
performs correction for systematics first and signal search thereafter, 
produces higher signal recovery rates on the average, while also being 
substantially faster than the simultaneous method. Independently of 
the method of search, once the signal is found, a far less time 
consuming full-fledged model, incorporating both the signal and systematics, 
must be employed to recover the correct signal shape. As a by-product 
of the tests on the K2 data, we find that for longer period sinusoidal 
signals the detection rate decreases (after an optimum value is reached) 
as the number of light curves used for systematics filtering increases. 
The decline of the detection rate is observable in both methods of filtering, 
albeit the simultaneous method performs better in the regime of relative high 
template number. We suspect that the observed phenomenon is linked to 
the increased role of low amplitude intrinsic stellar variability in the 
space-based data. This assumption is also supported by the substantially 
higher stability of the detection rates for transit signals against the 
increase of the template number.} 


\titlerunning{Periodic transit and variability search}
\authorrunning{Kov\'acs et al.}
   \maketitle
%

%
\section{Introduction}
The filtering of instrumental and environmental effects from
astronomical time series data is a vital ingredient in the data
processing pipelines for the time series collected by transiting
extrasolar planet (TEP) search projects.\footnote{Systematics are 
also important for planetary systems discovered by radial velocity 
technique, although here the most significant ones come from the 
star, attributed to various steller surface phenomena (see, e.g., 
Rajpaul et al.~\cite{rajpaul2015}).} The reason why systematics
filtering is so important in the search for TEPs is that these surveys
`indiscriminately' examine basically all stars (except perhaps obvious
giants) available in their photometric databases. Therefore, small 
systematics of the size of few percent or less become significant due 
to the leakage of their power (directly or via their harmonics) to 
the frequency range of interest and thereby blur the signal. 
Except for the aliasing, space-based surveys ({\it CoRoT} and 
{\it Kepler}) suffer from the same problem. 

Although there is a weak resemblance between today's systematics 
filtering methods and the widely used simple ensemble photometry 
(e.g., Honeycutt~\cite{honeycutt1992}), there is also a basic 
difference. In the case of ensemble photometry we do not fit the 
target light curve (LC) by time series carrying information on the 
systematics but rather assume that all stars in the field are affected 
by the same transparency change that can be accurately estimated by 
averaging out all incoming fluxes from the individual stars.\footnote{We may 
also apply color and spatial terms due to differential extinction. 
Even after these terms are fitted for, and applied, there remain 
systematics.} With the division of the target flux by the ensemble 
flux, we filter out systematic variations that are common to all 
stars, but any differences between sources (due to e.g. local 
topology, subpixel structure, etc.) are not corrected for. 

Modern methods of systematics filtering assume that the systematics 
are specific to each star, but they can be built up as the linear 
combinations of the systematics of other stars (or with the aid of 
other auxiliary measurable quantities, e.g., the width of the point 
spread function). The determination of the optimum linear combination 
is usually performed by standard least squares technique. 

This approach is drastically different from the one followed by the 
ensemble method, because here the LCs are `flattened out' as much as 
possible, which may lead to a substantial depression of the signal 
we are searching for. The success of the method depends on the relative 
degree of depression of the systematics and the signal. It is expected 
that in general, the signal `wins', since its properties are usually 
less common with those of the other LCs, whereas the contributing 
systematic effects are more likely shared also with other stars in 
the field.  

The method was implemented first in the Trend Filtering Algorithm (TFA, 
Kov\'acs, Bakos, \& Noyes~\cite{kovacs2005}) and in SysRem 
(Tamuz, Mazeh, \& Zucker~\cite{tamuz2005}). While TFA uses a 
`brute force' fit of many (several hundred) nearly randomly selected 
template stars from the target field, SysRem employs an iterative 
algorithm that results in a lower final number of correcting time series 
(albeit this number may not be well-defined in practice). This correcting 
set is built up from a large number of LCs of the target field, similarly 
to the popular Principal Component Analysis (PCA). Both of these methods 
can be run in `reconstructive' mode, with full (systematics$+$signal) 
model fit, once the signal frequency is found (and, in the case of SysRem, 
once the basis of trends has been determined). This results in a much 
better reproduction of the original signal shape and cures the depression 
caused in the signal search phase. 

In spite of the ability of reconstructing the signal after its initial 
detection, efforts have been made to decrease the level of signal 
depression and minimize the chance of possible signal loss during the 
signal search phase of the analysis. This has usually been tackled by 
a careful selection of the template time series, keeping their number to 
a minimum and introducing more general probabilistic (i.e., Bayesian) 
treatment of the problem. Kim et al.~(\cite{kim2009}) use a 
hierarchical clustering method to select an optimum number of 
co-trending LCs. Chang, Byun, \& Hartman~(\cite{chang2015}) adopt the 
same method as part of their pipeline for precise cluster photometry. 
A PCA-based  criterion is used in the algorithm proposed by Petigura \& 
Marcy~(\cite{petigura2012}) for the analysis of the {\it Kepler} 
LCs. The more involved, PDC-MAP pipeline (Stumpe et al. \cite{smith2012}; 
Smith et al. \cite{smith2012}) of the {\it Kepler} mission also utilizes 
PCA for selecting the basis vectors for the systematics correction. In a 
similar manner, Roberts et al.~(\cite{roberts2013}) discuss the advantage 
of using Bayesian linear regression for robust filtering and employ entropy 
criterion for selecting the most relevant corrections while maintaining the 
original signal as much as possible.  

The method has also been extended by including additional effects, such 
as stellar variability (Alapini \& Aigrain~\cite{alapini2009}, Kov\'acs \& 
Bakos~\cite{kovacs2008}) and simultaneously considering multiplicative and 
additive systematics, both of which may be relevant when a wide  
brightness range is considered (e.g., in the {\it CoRoT} data, see the 
SARS algorithm by Ofir et al.~\cite{ofir2010}).   

The K2 mission (the successful continuation of the {\it Kepler} mission 
using two reaction wheels -- see Howell et al.~\cite{howell2014}) inspired 
further ideas to revisit the problem of trend filtering.\footnote{Here, 
admittedly somewhat loosely, we use the term `trend filtering' as a synonym 
for `systematics filtering'. Smith et al.~(\cite{smith2012}) make 
distinction between co-trending and de-trending, with the latter reserved 
for high-pass filtering, without respect to the origin of the trend.} 
Several papers have been published on this subject, focusing mainly on the 
wobble of the spacecraft due to the periodic ignition of the thrusters, 
to stabilize the pointing. Vanderburg \& Johnson~(\cite{vanderburg2014}) 
employed a simple, yet effective method to correct for effects 
due to the wobble. Their method utilizes the non-linear correlation 
between the pixel position and flux variation, and leads to a substantial 
(a factor of $2-5$) improvement in the RMS of the light curves. The method 
of Vanderburg \& Johnson~(\cite{vanderburg2014}) is a specific extension 
of the method of light curve correction based on image properties 
(external parameter decorrelation, EPD, see Bakos et al.~\cite{bakos2010} 
for ground-based wide-field surveys, and those of, e.g., Knutson et al.~\cite{knutson2008} 
and Cubillos et al.~\cite{cubillos2014} for the treatment of the Spitzer 
planet transit and occultation data). In a very recent paper by 
Huang et al.~(\cite{huang2015}) the authors employ a combination of 
various filtering techniques (including EPD, TFA and Fourier) and, in 
the most favorable brightness range, reach a precision of $15$--$20$~ppm 
on a $6.5$~hours time span -- i.e., hitting the precision of the 
original {\it Kepler} mission.   

Focusing also on the roll angle variation in the K2 data, 
Aigrain et al.~(\cite{aigrain2015}) developed a method based on the 
simultaneous modeling of this particular systematics and intrinsic stellar 
variability within the framework of Gaussian processes. This allows for a 
great flexibility in determining the target-dependent effect of the roll 
angle variation, together with the conservation of the intrinsic signal. 
The authors report broad agreement in the gain of precision with the one 
obtained by Vanderburg \& Johnson~(\cite{vanderburg2014}). 

Foreman-Mackey et al.~(\cite{foreman2015}) do not focus on the roll angle 
variation alone but use a relative large number of eigen LCs from their 
PCA to model the K2 LCs as the linear combination of these eigen 
LCs and a transit model. Similarly to the approach of 
Aigrain et al.~(\cite{aigrain2015}), this is also a full time series 
modeling in the sense that they conduct a simultaneous search for the 
best-fitting combination of systematics and signals. For the computationally 
challenging task of searching for the transit parameters while finding 
also the true contribution of the systematics, 
Foreman-Mackey et al.~(\cite{foreman2015}) opt for the maximum likelihood 
approach in which it is assumed that the transit events are independent. 
With this assumption the joint distribution function can be computed as a 
product of the likelihoods of the individual events. This observation 
allows a considerable speed up of the otherwise slow minimization process, 
since a part of the computation can be performed on a coarse parameter 
grid that can be used thereafter for the fine grid search by simply 
interpolating on the coarse grid. Although the authors do not deal with 
the precision of their finally derived LCs, from their Figure~2 and 
Table~1 of Vanderburg \& Johnson~(\cite{vanderburg2014}) it seems that 
they both reach similar precision of $\sim 33$~ppm on an integration time 
span of $6$~hours. This is only $50$~\% larger than the corresponding 
figure for the original {\it Kepler} mission in the same brightness range 
of $11-12$~mag. In a subsequent paper the method is extended to general 
variability search by using a Fourier, instead of a boxcar representation 
of the signal (Angus et al.~\cite{angus2015}).  

Yet another recent work by Wang et al.~(\cite{wang2015}) introduces 
a different method aiming at simultaneous systematics filtering and signal 
preservation. The method (Causal Pixel Model) is based on the autoregressive 
and co-trended maximum likelihood estimate of each pixel flux associated 
with a given target. Each time series value is predicted from a fit 
computed by the omission of the values close in time to that observation.
The size of the window of omission is chosen freely based on the expected 
duration of the transit event. By construction, the resulting filter is 
transit signal preserving (but, because of the autoregressive fit, it is 
also a stellar variability `killer'). The authors report a consistently 
better performance of their method when compared to the standard Kepler 
pipeline PDC. 

The idea of the full-fledged (i.e., systematics$+$signal) search has 
also come up in the context of ground-based surveys. To increase of the 
transit detection capability of the MEarth project, 
Berta et al.~(\cite{berta2012}) employ a nightly full model fit to the 
target LCs and then combine the individual likelihood functions into a 
joint likelihood. The solution found with this probabilistic approach 
helps to avoid nightly overfitting, which is a common problem in 
multi-parametric fits of small datasets.  

Stimulated by the above efforts, this paper deals with the performance 
of the full-fledged frequency search. Naturally, the full model approach 
is preferable once the signal period is found (Kov\'acs et al.~\cite{kovacs2005}). 
Also, by fitting systematics only, the underlying signal suffers from 
some level of depression, which jeopardizes the detection capability. 
However, it is unclear if in the {\em period search phase} the increased 
degree of freedom due to the inclusion of the signal model in the 
systematics fit will not increase the false alarm rate. Last but not 
least, for signal search, the full model approach is much more intensive 
computationally than the partial model fit (i.e., assuming no signal, 
fitting only the systematics and performing the signal search on the 
residuals thereafter). Therefore, it is important to investigate if the 
full model search is worth the effort (i.e., if the increased complexity 
and execution time is compensated by the increased detection efficiency).

%
\section{Simultaneous fit for systematics and periodic signal}
Let us assume that the photometric pipeline supplies a large set 
of time series $\{x(j,i); j=1,2,...,L; i=1,2,...,N \}$, containing 
$L$ light curves, each with $N$ data points, and, for simplicity, 
sampled on the same timebase $\{t(i); i=1,2,...,N \}$. Following 
the TFA methodology (for simplicity) we model each of the observed 
time series as a linear combination of suitably chosen $M$ cotrending 
light curves and a Fourier sum of arbitrary order $K/2$. For any 
target time series $\{y(i); i=1,2,...,N \}$ (selected from $\{x(j,i)\}$) 
at any given trial frequency $\nu$ we minimize the residuals ${\cal D}$ 
between the model and data following the standard least squares (LS) 
approach 
%
%
\begin{equation}
{\cal D} = \sum_{i=1}^N \left(y(i)-\sum_{j=1}^M a_{\rm j}S(j,i) - \sum_{j=1}^{K+1} b_{\rm j}F(j,i)\right)^2 
\hskip 2mm , 
\end{equation}
where $\{S(j,i); j=1,2,...,M; i=1,2,...,N \}$ denote the set of cotrending 
time series and $\{F(j,i); j=1,2,...,K+1; i=1,2,...,N \}$ stand for the Fourier 
representation of the underlying signal. In particular, at any given instant  
of time $t(i)$, $F(1,i)=1, F(2,i)=\sin(\varphi), 
F(3,i)=\cos(\varphi), ..., F(K,i)=\sin(K\varphi), 
F(K+1,i)=\cos(K\varphi)$, with phase $\varphi=2\pi\,\nu\,t(i)$ 
(and, of course, the reference epoch for $\{t(i)\}$ is arbitrary). 

\begin{figure}
 \vspace{0pt}
 \includegraphics[angle=-90,width=85mm]{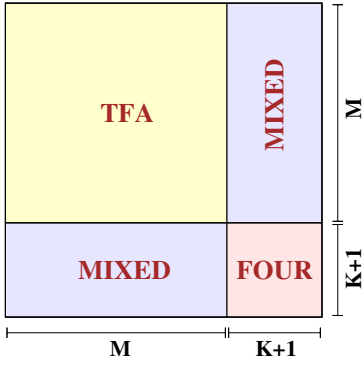}
 \caption{Main blocks of the normal matrix of the full-fledged systematics 
          filtering least squares problem (see Eqs.~(1), (2)). In the 
	  course of the period search, only the MIXED and FOUR blocks 
	  change (the latter containing the pure Fourier terms of the fit).}
\label{ls_matrix}
\end{figure}

The LS condition above leads to normal matrix schematically shown in 
Fig.~\ref{ls_matrix}. We see that when scanning the various test frequencies, 
the TFA part of the matrix (typically the dominant part of the full 
matrix) does not change. This block structure of the normal matrix allows 
us to ease the otherwise very heavy computational load required by 
the re-computation and inversion of the normal matrix each time a new 
frequency is tested. 

%

In detail, the following set of linear equations is obtained when satisfying 
the LS condition posed by Eq.~(1)
%
%
\begin{equation}
\left( \begin{array}{lr}
TF & MI^T \\
MI & FO \end{array} \right)
\hskip 0mm 
\left( \begin{array}{cc}
a \\
b \end{array} \right) 
\hskip 0mm = 
\left( \begin{array}{cc}
YT \\
YF \end{array} \right) 
\hskip 2mm ,
\end{equation}
where $\{a\}$ and $\{b\}$ are, respectively, $M$- and 
$K+1$-dimensional column vectors. The block elements of the normal matrix 
and the right-hand-side vectors are computed as follows:  

%
\begin{eqnarray}
TF(i,j)   & = & \sum_{k=1}^N S(i,k)S(j,k) \hskip 1mm , 1\leq i \leq M,   1\leq j \leq M   \nonumber \\
MI(i,j)   & = & \sum_{k=1}^N F(i,k)S(j,k) \hskip 1mm , 1\leq i \leq K+1, 1\leq j \leq M   \nonumber \\
MI^T(i,j) & = & \sum_{k=1}^N S(i,k)F(j,k) \hskip 1mm , 1\leq i \leq M,   1\leq j \leq K+1 \nonumber \\
FO(i,j)   & = & \sum_{k=1}^N F(i,k)F(j,k) \hskip 1mm , 1\leq i \leq K+1, 1\leq j \leq K+1 \nonumber \\
YT(i)     & = & \sum_{k=1}^N S(i,k)y(k)   \hskip 1mm , 1\leq i \leq M   \nonumber \\
YF(i)     & = & \sum_{k=1}^N F(i,k)y(k)   \hskip 1mm , 1\leq i \leq K+1 \hskip 2mm .
\end{eqnarray}
In solving Eq.~(2) first we observe that the cotrending (TFA) part needs to 
be computed only once, since this is constant during the frequency scan. 
Furthermore, if we opt for solving the system with matrix inversion, we 
can utilize the formula valid for block matrix inversion, that may also 
give some advantage in decreasing the computational load. Further speed-up can 
be gained by utilizing the positive definitive nature of the normal matrix and  
use, e.g., Cholesky factorization, or the more general QR decomposition 
(e.g., Press et al.~\cite{press1992}). The inverse of the normal matrix 
of Eq.~(2) constitutes the following blocks (the Helmert-Wolf blocking, 
see the proof by Banachiewicz~\cite{banachiewicz1937} and applications, 
e.g., by Rajan \& Mathew~\cite{rajan2012}):

%
\begin{eqnarray}
\left( \begin{array}{lr}
TF & MI^T \\
MI & FO \end{array} \right)^{-1} 
= 
\left( \begin{array}{lr}
TF^{-1}+G*MI*TF^{-1} & -G \\
-SC^{-1}*MI*TF^{-1} &  SC^{-1} \end{array} \right) \hskip 2mm . 
\end{eqnarray}
In the above eqations symbol $*$ denotes matrix multiplication, 
$SC$ is the Schur complement of the normal matrix ($SC=FO-MI*TF^{-1}*MI^{T}$) 
and $G=TF^{-1}*MI^T*SC^{-1}$. Although employing these formulae for the 
inversion results in an increase of $20$--$30$\% in the speed of execution 
(relative to the simple Gaussian inversion of the full matrix), the 
full-fledged systematics filtering is still very expensive, due to the 
large size of matrix $TF$, carrying the information on systematics. 
Our experience shows that for a moderate size time series 
(e.g., with few thousand data points) and $200$ TFA template time 
series the full-fledged (hereafter {\sc tfadft}) search is $2$--$3$-times 
slower than the ``filter first, search for signals thereafter''-method 
(hereafter {\sc tfa+dft}). With doubling the TFA template size this 
ratio increases to nearly $10$.   
 
Within the above framework we can substitute the model functions 
$\{F(j,i)\}$ with any other, frequency-dependent functions, better 
representing the signal to be searched for. We can also allow 
for additional non-linear parameters in the model functions, further 
increasing the complexity of the problem (see Foreman-Mackey~et. al~\cite{foreman2015} 
for boxcar signals). Although Fourier representation is sub-optimal 
for transit signals (Kov\'acs, Zucker, \& Mazeh~\cite{kovacs2002}), 
for the purpose of this paper it is suitable, since both the {\sc tfadft} 
and the {\sc tfa+dft} models are tested within the same framework. 
Additional applications of the Fourier base for transit search can 
be found in Moutou et al.~(\cite{moutou2005}) and 
Samsing~(\cite{samsing2015}).

%
\section{Comparison of the performances of the time series models}
Here we report on a detailed numerical testing of the full time 
series model fit ({\sc tfadft}) and the two-step partial fit 
({\sc tfa+dft}). Our testing ground is the frequency spectra of the 
various test signals. These frequency spectra are based on the RMS 
of the residuals of the fitted data (see Eq.~(1)). To retain the 
more traditional pattern of the frequency spectra, the residual 
spectra are obtained from {\cal D} as follows

%
\begin{eqnarray}
P(\nu) = {\sqrt{\cal D}-\sqrt{\cal D_{\rm max}} \over \sqrt{\cal D_{\rm min}}-\sqrt{\cal D_{\rm max}}} \hskip 2mm , 
\end{eqnarray}
where the indices refer to the max/min values of {\cal D}. The function 
$P(\nu)$ is akin to the amplitude spectrum and it is obviously confined 
to $[0,1]$. The characterization of the signal-to-noise ratio of the 
highest peak in $P(\nu)$ goes in the standard way

%
\begin{eqnarray}
{\rm SNR} = {P(\nu_0)-{<}\,P(\nu)\,{>} \over \sigma(P(\nu))} \hskip 2mm . 
\end{eqnarray}
Here $P(\nu_0)$, ${<}\, P(\nu)\, {>}$ and $\sigma(P(\nu))$ denote, 
respectively, the power at the peak frequency, the average of the power 
over the waveband where the SNR is referred to and the standard deviation 
of the `grass' (i.e., the noise) component of the spectrum. This latter 
phrase means that we omit `outliers' (i.e., high peaks) when 
$\sigma(P(\nu))$ is computed.\footnote{We employ iterative $3\sigma$ 
clipping in finding the RMS of the noise component of the spectrum.} 
Please note that with the above definition SNR can be a sensitive function 
of the waveband of reference, especially if the noise is colored. 

In the following we perform a two-step analysis of the efficiency of 
the {\sc tfadft} and {\sc tfa+dft} approaches to systematics filtering. 
In the first step we examine the signal recovery properties of these 
methods on a simple two-component time series. Then, using 
subsets of the databases of the HATNet\footnote{http://hatnet.org/} 
and K2\footnote{http://keplerscience.arc.nasa.gov/K2/} projects, 
we inject various signals in the observed data and check the discovery 
rates for the two methods. 

%
\subsection{Two-component signal test}
We start with the simplest possible scenario of two noisy sine functions 
with known parameters.  

%
\begin{eqnarray}
y(i) & = & g(i) + A_2\sin(2\pi\nu_{2}t_{\rm i})  + gn2(i) \nonumber \\
g(i) & = & A_1\sin(2\pi\nu_{1}t_{\rm i}) + gn1(i) \hskip 2mm . 
\end{eqnarray}
Here $gn1$ and $gn2$ denote independent Gaussian white noise. Function 
$g$ is considered as being the contribution from the systematics, whereas 
the second sine component constitutes the signal we are searching for. 
In this basic test we make function $g$ known for the routines. Under 
this circumstance, {\sc tfadft} should yield an exact match to the data, if 
the signal is noiseless ($gn2=0$).\footnote{Note that in this setting $g$ 
can be arbitrary. We take it as a noisy sine function only for simplicity 
and emphasize the close similarity between the problem discussed in this 
paper and the commonly used pre-whitening technique for various astronomical 
signals. Note also that having independent noise in $g$ prevents singularity 
when the full model is used.} Therefore, the more interesting question is 
what happens if we increase the noise component of the signal. We examine 
this question for $100$ signal frequencies by uniformly scanning the 
$[0.4,0.5]$d$^{-1}$ frequency range. Other time series parameters are 
fixed to the following values: $A_1=0.02$, $\nu_{1}=0.45$, $\sigma_1=0.001$, 
$A_2=0.01$ and $\sigma_2=0.06$ for the low- and $\sigma_2=0.0001$ for the 
high-SNR case. For each signal frequency we add the same realization of 
the noise to the signal to keep track only of the effect of changing 
signal frequency. For sensing the effect of noise realization, we repeat 
the frequency scan for $10$ different realizations. All tests are performed 
on the timebase of one of the light curves from the HATNet field to be 
described in Sect.~3.2.    

%
\begin{figure}
 \vspace{0pt}
 \includegraphics[angle=-90,width=85mm]{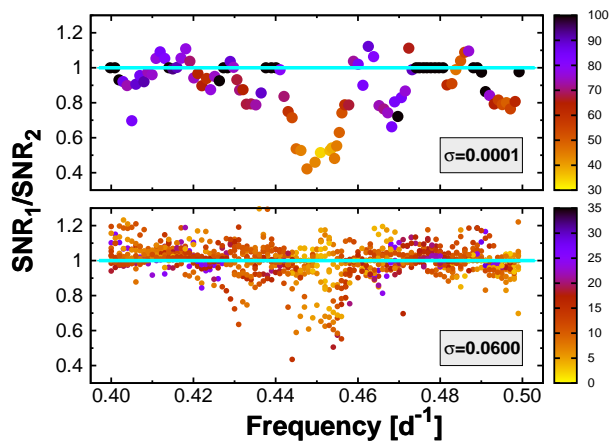}
 \caption{Variation of the SNR ratios of the {\sc tfa+dft} (SNR$_1$) 
 and {\sc tfadft} (SNR$_2$) spectra as a function of the injected 
 signal frequency for the two-component time series described in the
 text. The shading is proportional to SNR$_2$ as indicated by 
 the sidebars. For realistic (noisy) time series, the signal detection 
 capability of the two methods become nearly the same. The result 
 shown in the bottom panel is based on $10$ realizations of the Gaussian 
 noise added to the time series.}
\label{snr-2sin}
\end{figure}

%
\begin{figure}
 \vspace{0pt}
 \includegraphics[angle=-90,width=85mm]{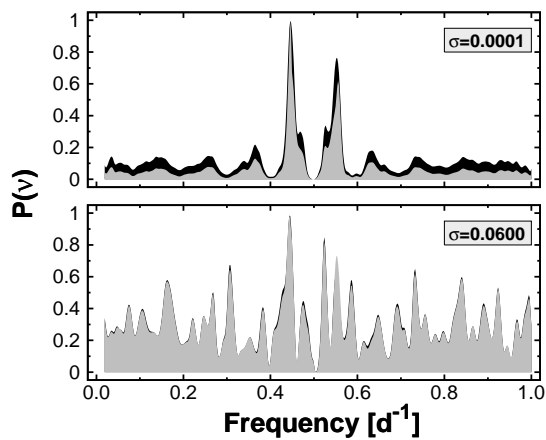}
 \caption{Illustration of the dependence of the frequency spectra on 
        noise level (upper vs lower panels) and on the method of data 
	analysis (lighter shade: full time series model; darker 
	shade: partial time series model). In this example the 
	frequency of the test signal is $0.446$d$^{-1}$, very close 
	to that of the systematics. Farther away from the frequency 
	of the systematics the difference between the two methods is 
	much smaller, even in the noiseless case.}
\label{sp-2sin}
\end{figure}

The dependence of the SNR of the frequency spectra on the signal frequency 
with the above fixed component of systematics is shown in Fig.~\ref{snr-2sin}. 
As expected, the high-SNR case (upper panel) clearly shows the advantage 
of the full model fit over the partial one in the immediate neighborhood of 
the frequency of the systematics. However, it is important to note that 
in this high-SNR range the better performance of {\sc tfadft} does not have 
much use, since in both cases the signal is detected with very high 
significance, even in the close neighborhood\footnote{That is, less than 
the characteristic FWHM/2 of $\sim {1 \over 2}T^{-1}$, which is $0.01$d$^{-1}$ 
in this case.} of the frequency of the systematics. For low SNR the situation 
changes, and the difference between the two models diminishes, as shown in 
the bottom panel of Fig.~\ref{snr-2sin}. Although there is still some signal 
frequency dependence (affected by noise realization), most of the SNR values 
are equal within $\sim 10$\%, with a slight preference toward the partial 
model. 

For a more straightforward illustration of the convergence of the two 
signal search methods as the noise is increased, in Fig.~\ref{sp-2sin} 
we display the frequency spectra related to the test with the signal 
frequency $0.446$d$^{-1}$ (very close to $0.45$d$^{-1}$, the frequency 
of the systematics, which is the regime where {\sc tfadft} is 
expected to provide the greatest performance enhancement over {\sc tfa+dft}). 
From the figure it is clear that {\sc tfadft} outperforms {\sc tfa+dft} 
when the noise level is very low (when, however, the detection 
is not a problem for either of the methods). On the other hand, at a 
more reasonable noise level the two methods behave comparably. 

We conclude from this simple test (clearly favorable for the full time 
series model) that, in general, from pure signal detection point of view, 
the full model does not necessarily have advantage over the partial one 
that may justify its use. In specific cases of nearly coinciding signal 
and systematics frequencies we may gain some advantage from the full 
model, but it is unclear if the slightly larger detection power is worth  
the additional computational burden (especially in more realistic cases 
with high TFA template numbers - see Sect.~3.3).

%
\subsection{Ensemble test: datasets and signals}
In the previous section we used test data that were purely artificial, 
with known systematics and signal. However, for real observations 
we do not know the exact form of the systematics. 
In this case we build up (if possible) a simple model for the systematics 
and use this as an approximation of the true systematics. In practice, 
however inexact this approach is, it usually leads to very impressive 
improvement in signal detection rates. In the present context the inexact 
model for the systematics acts as an unknown noise component, influencing 
also the `full' time series models and making them more similar to 
the partial models. Whether this `less partial' model is able to outperform 
the `fully partial' model (based on separate systematics and signal fits) 
depends on the fraction of the unknown constituents in the adopted model 
of systematics and other factors, e.g., the spectral behavior of the time 
series at frequencies different from that of the signal. 

\begin{figure}
 \vspace{0pt}
 \includegraphics[angle=-90,width=85mm]{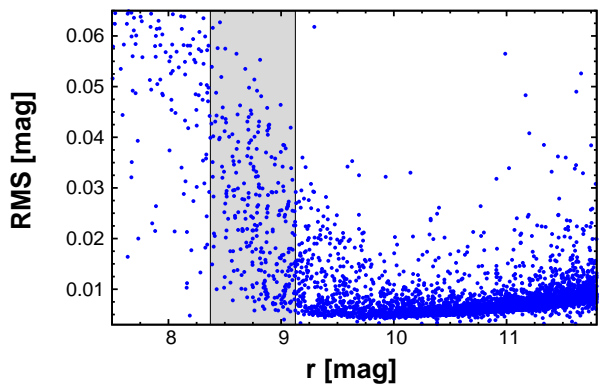}
 \caption{Magnitude-RMS plot for HATNet field \#317 (TFA-filtered 
 data with high TFA template number of $990$ are plotted). Shaded 
 area shows the region of the $300$ stars we choose for the tests 
 presented in this paper.}
\label{mag_sig}
\end{figure}

We inject various signals into sub-samples of light curves 
observed by HATNet (Bakos et al.~\cite{bakos2004}) and by the Kepler 
two-wheel mission during the phase of Campaign 1 
(K2, Howell et al.~\cite{howell2014}). The light curves from the HATNet 
database have properties that are typical for HATNet observations:  
they have been collected from a $10^{\circ}\times10^{\circ}$ field, 
containing nearly $39000$ objects from $\sim 6$~mag to $\sim 14$~mag 
(in Sloan {\em r}-band). Each object has a photometric time series with 
up to $8600$ datapoints spanning $165$ days. For economical testing we 
choose only the first $2000$ data points from each time series -- this  
cut decreases the time span down to $51$ days. We limit our 
study to $300$ objects at the brighter magnitude end between $8.4$ 
and $9.1$~mag. Most of these objects suffer from saturated pixels, 
due to their bright magnitudes (see Fig.~\ref{mag_sig}). Because of 
their large systematics, we select these objects to sense the 
differences between the methods tested more effectively. We choose 
the EPD (External Parameter Deconvolved) LCs -- see 
Bakos et. al.~(\cite{bakos2010}). Although these LCs are largely free 
from strong systematics, TFA introduces further improvements.  
 
For a quick and easy access to the light curves of Campaign~1 of 
the K2 mission, we resort to the depository of the K2 HAT Light Curve 
project\footnote{http://k2.hatsurveys.org/archive/} as described by 
Huang et al.~(\cite{huang2015}). To remain statistically compatible with 
the sample size used in the test of the HATNet data, we select a sample 
of $300$ stars from their {\sc object-summary.csv} file. All these stars 
have UCAC4 identifications\footnote{http://cdsarc.u-strasbg.fr/viz-bin/Cat?I/322A} 
with Johnson $V$ magnitudes, which we use to constrain the sample between 
$V=9.59$--$10.44$~mag. Unlike Huang et al.~(\cite{huang2015}), we include 
objects from all channels, both in the sample of these $300$ stars and 
also in the selected TFA templates (this enables us to use larger template 
numbers). For the templates we restrict the selection to stars brighter 
than $V=11.5$~mag. As for the HATNet data, we constrain the original number 
of data points per LC of $3820$ to $2000$. This cut decreses the time span 
of the data from $82$~days to $44$~days. At all template numbers the templates 
are distributed on a nearly uniform grid in the full field of view (following 
the original idea of template selection as described in 
Kov\'acs et al.~\cite{kovacs2005}). We use the {\sc best aperture light curves} 
and select the direct photometric values (column \#5 in the light curve files, 
annotated as IM\#\#, where \#\# denotes the aperture size). Although using 
these direct photometric values have the advantage of performing our injected 
signal tests as `purely' as possible, it is sub-optimal in terms of the 
accessible signal sensitivity, since it lacks EPD correction, which carries 
away a great part of the systematics in the K2 data. We note that for the 
HATNet data our choice of the EPD magnitudes is justified on the basis of 
the large variation of the original photometric fluxes, which makes outlier 
handling more cumbersome in the TFA analysis (in the case of the K2 data the 
straight photometric fluxes behave in a way that is more easy to tackle).

%

Two basic signal types are tested. The first one is a simple 
sinusoidal signal whereas the second one is a periodic transit signal. 
As shown in Table~\ref{test-signals}, the combination of the periods, 
amplitudes (transit depths) are different for the seven test signals.  
In all cases the amplitudes are selected to be low enough to avoid 
detection in the non-filtered (original) data but high enough to allow 
detection in the systematics-filtered data. Periods are chosen to be 
short enough to avoid data sampling issues in the case of the the 
transit signal when testing the HATNet data. Signals \#1 and \#2 are 
used for testing the HATNet data, whereas \#3 -- \#7 are employed on 
the K2 data. For the latter, \#3 and \#4 are akin to \#1 and \#2, 
except for their amplitudes, that are adjusted to the considerable 
lower noise level of the K2 data. Signal \#5 is intended to demonstrate 
the stabilization of the detection rate for sinusoidal signals in 
the K2 data by the decrease of the period. The longer period sinusoidal 
signal \#6 (along with the somewhat shorter period signal \#3) is 
for showing the impaired detection rate for longer period sinusoidal 
signals on the K2 data at higher TFA template numbers. Signal \#7 is 
devoted to show the opposite behavior of the transit signals on the 
same dataset even at these longer periods.  
 
\begin{figure}
 \vspace{0pt}
 \includegraphics[angle=-90,width=85mm]{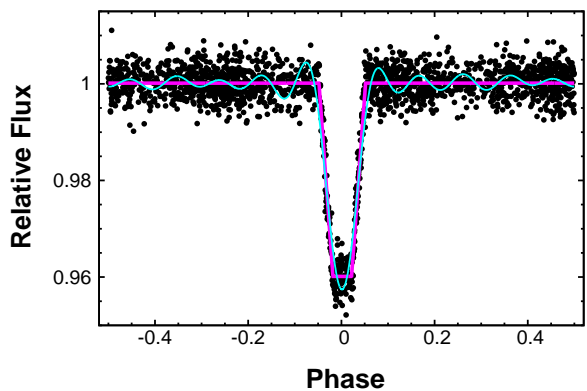}
 \caption{Illustration of the Fourier approximation of the 
          transit-like signal \#2. The synthetic trapezoidal signal 
	  is shown by pink line. Black dots show the synthetic 
	  signal after adding a Gaussian noise with $\sigma=0.003$. 
	  The result of a 10-th order Fourier fit to these noisy data 
	  is shown by light blue line.}
\label{transit-four}
\end{figure}

As mentioned, we opted for using a Fourier-based representation of 
the signals to be tested, in spite of the poor capability of the Fourier 
decomposition in the case of more appropriate (i.e., short event 
duration) transit signals. Furthermore, within the framework of standard 
LS fitting, with the increasing order of the Fourier sum, it becomes 
more vulnerable against data gaps and outlying data points. Therefore, 
by choosing a long transit length (relative to the period), more 
representative for short period systems, we are able to use a reasonably 
low-order Fourier sum that is stable but still yields an acceptable 
representation of the transit shape (see Fig.~\ref{transit-four}).

%

%
\begin{table}
 \centering
 \begin{minipage}{140mm}
  \caption{Test signal parameters}
  \label{test-signals}
  \scalebox{1.00}{
  \begin{tabular}{clccc}
  \hline
   Name  & Type & Ampl. & Freq. & $Q_{\rm tran}$ \\
 \hline
 $\#1$ &  sine    & $0.0100$ & $0.24$ &    $-$ \\
 $\#2$ &  transit & $0.0400$ & $0.64$ & $0.10$ \\
 $\#3$ &  sine    & $0.0002$ & $0.24$ &    $-$ \\
 $\#4$ &  transit & $0.0005$ & $0.64$ & $0.10$ \\
 $\#5$ &  sine    & $0.0001$ & $0.64$ &    $-$ \\
 $\#6$ &  sine    & $0.0001$ & $0.21$ &    $-$ \\
 $\#7$ &  transit & $0.0005$ & $0.21$ & $0.10$ \\
\hline
\end{tabular}}
\end{minipage}
\begin{flushleft}
{\bf Notes:} Amplitudes/transit depths and frequencies are given 
in relative fluxes and [$d^{-1}$], respectively. The ratio of the total 
transit time to the period is denoted by $Q_{\rm tran}$. The phase of 
the transit is chosen to yield near the average expected number 
(i.e., Q$_{\rm tran}\times N=200$) of in-transit data points for 
the HATNet data. Signals \#3 -- \#7 are used for testing the K2 data 
(see Sect.~3.4).  
\end{flushleft}
\end{table}

%
\subsection{Ensemble test: results based on HATNet}
By injecting known test signals in the target light curves we are able 
to check if the signal is detectable in the LS Fourier spectra. Instead 
of giving only a lower SNR limit as the sole detection criterion, which 
must be employed for unknown signals, here we utilize the fact that the 
signal frequency is known. Therefore, we extend the detection criterion 
by a condition on the peak frequency. At any given SNR level we consider 
the test signal detected if the peak frequency is within the 
$\pm\Delta\nu$ interval of the known injected frequency. We choose 
$\Delta\nu=0.01$, allowing near FWHM mismatch as an upper limit. For 
simplicity we do not consider alias detections. For a given SNR$_{\rm min}$ 
we define the detection ratio DR as the ratio of the number of detections 
with SNR$>$SNR$_{\rm min}$ to the total number of objects tested. All 
time series are analyzed in the $[0,1]$d$^{-1}$ interval and the SNR 
values refer to this frequency band.  

\begin{figure}
 \vspace{0pt}
 \includegraphics[angle=-90,width=85mm]{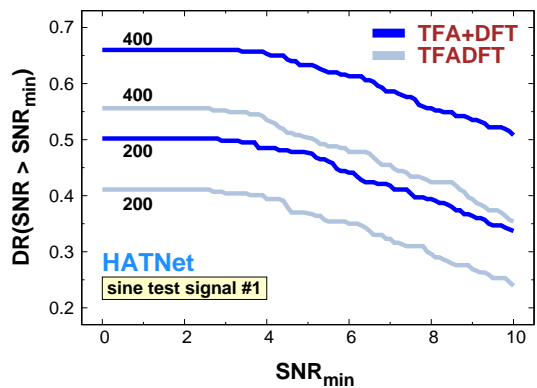}
 \caption{Detection ratio as a function of SNR for the sine test 
          signal \#1 of Table~\ref{test-signals} on the HATNet 
	  sub-dataset. DR denotes the relative number of detections 
	  with SNR$>$SNR$_{\rm min}$. The curves are labelled by the 
	  TFA template numbers. The partial time series models 
	  (blue lines with label {\sc tfa+dft}) yield significantly 
	  higher detection rates than the full models (fainter lines with 
	  label {\sc tfadft}).}
\label{sine-dr}
\end{figure}

The result for the sinusoidal test signal is shown in Fig.~\ref{sine-dr}. 
For sensing the dependence of the algorithm performance on the TFA 
filter size, we executed two runs with template numbers of $200$ and $400$. 
{\em The partial time series model {\sc tfa+dft} clearly outperforms the 
full model {\sc tfadft} in both cases} (tests with other template numbers 
show that this trend continues both for low and high template numbers). 
Assuming that the test signal frequency was unknown, at a more secure 
detection regime, say with SNR$>8$ and TFA template number of $400$, 
{\sc tfa+dft} would have been able to detect some $55$\% of the injected 
signals. For {\sc tfadft} this detection ratio is $43$\%. 

%

%
\begin{figure}
 \vspace{0pt}
 \includegraphics[angle=-90,width=85mm]{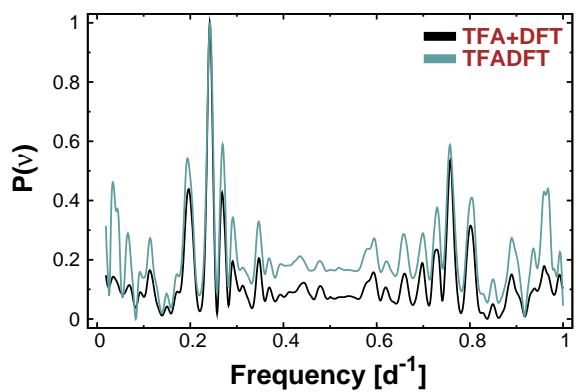}
 \caption{Example of the better performance of the partial time 
          series model {\sc tfa+dft} over the full model {\sc tfadft} 
	  for the sine test signal (test signal \#1 in 
	  Table~\ref{test-signals}). The spectra of one of the members 
	  of the $300$ objects used in this section for testing  
	  the HATNet database are shown. The number of TFA templates is 
	  equal to $400$. Both spectra are normalized to $1$ at the highest 
	  peak.}
\label{sine-sp}
\end{figure}

%
\begin{figure}
 \vspace{0pt}
 \includegraphics[angle=-90,width=85mm]{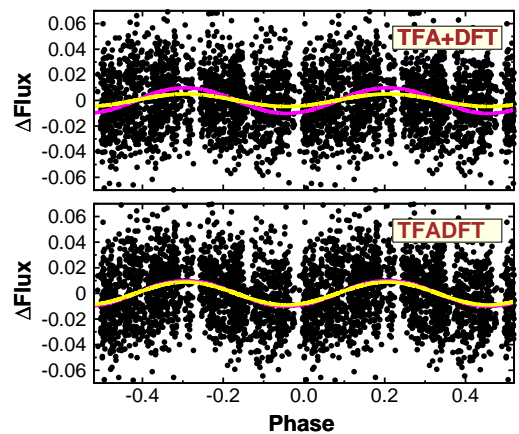}
 \caption{Folded light curves for the object with the frequency 
          spectra shown in Fig.~\ref{sine-sp}. The input sine signal 
	  is shown by pink line. The best single-component Fourier 
	  fit is overpolotted by yellow line. For better visibility, 
	  the data are plotted with twice of the test period. The 
	  partial time series model (upper panel) yields significantly 
	  lower amplitude fit in spite of finding the signal with a  
	  considerably higher significance than the full model search.}
\label{sine-lc}
\end{figure}

We exhibit the better detection capability of the partial model on one 
of the members of the testing set. Fig.~\ref{sine-sp} shows the frequency 
spectra for the two types of analysis. The higher variation of the 
spectrum away from the injected frequency in the case of full 
model results in an SNR of $9$, compared to $20$ of the partial model. 
Although the case displayed is characteristic of the more extreme examples, 
the SNR values of the full models rarely reach those of the partial models. 
This leads to the final accumulation of the higher detection ratio for 
the partial model, as shown in Fig.~\ref{sine-dr}. 

It is also interesting to examine the filtered time series resulting 
from the partial and full models. The folded time series, together 
with the input synthetic (noiseless) signals for the target above are 
shown in Fig.~\ref{sine-lc}. We see that the application of the partial 
model leads to a substantial ($\sim 50$\%) squeezing of the signal. 
This is because the partial model is based on the premise of the absence 
of an underlying signal and this may lead to fitting any variation in 
the time series that has some correlation with the TFA template set. 
For the same reason, the partial model treats true systematics better 
than the full model, because in the latter the flexible Fourier part 
of the fit may select the systematics as a real signal, when the wrong 
frequency is tested, thereby decreasing the fitting power of the TFA 
templates.\footnote{In the parlance of the Bayesian framework, the 
procedure of first filtering systematics from the light curve, then 
searching for periodic signals, and finally carrying out a combined 
fit, can be understood as placing a prior constraint on the frequency 
of the signal component of the model (i.e., that it must be close to 
that found when the systematics filtering and period search are carried 
out separately). This effectively reduces the likelihood of signal 
frequencies where the model for the systematics also has significant 
power, increasing the sensitivity to low-amplitude signals at other 
frequencies.} 

%
\begin{figure}
 \vspace{0pt}
 \includegraphics[angle=-90,width=85mm]{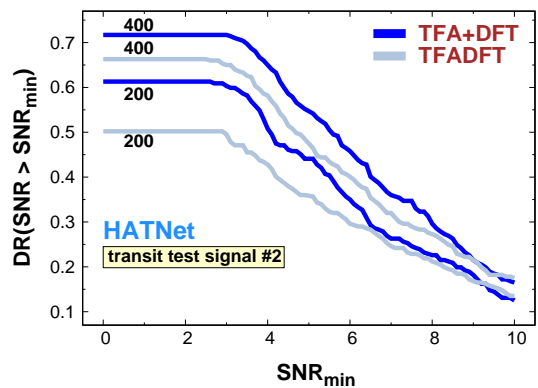}
 \caption{Detection ratio as a function of SNR for the transit test 
          function (test signal \#2 in Table~\ref{test-signals}). 
	  Notation is the same as in Fig.~\ref{sine-dr}. The better 
	  detection capability of {\sc tfa+dft} over {\sc tfadft} 
	  is observable here too, albeit the effect is considerably 
	  smaller than for the sine test signal case.}
\label{transit-dr}
\end{figure}

\begin{figure}
 \vspace{0pt}
 \includegraphics[angle=-90,width=85mm]{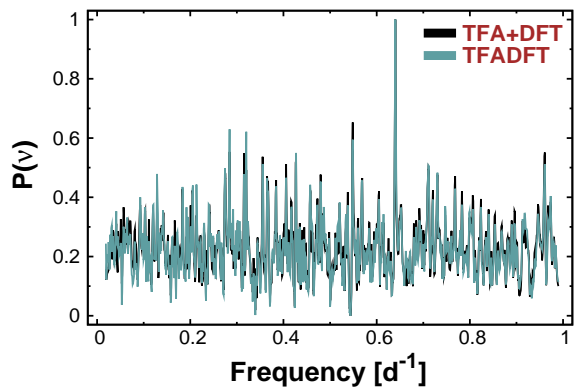}
 \caption{Example of the similar performance of the partial time 
          series model {\sc tfa+dft} and the full model {\sc tfadft} 
	  for the transit-type test signal (test signal \#2 in 
	  Table~\ref{test-signals}). A $10$-th order Fourier sum 
	  is used to search for the transit signal. Notation is the 
	  same as in Fig.~\ref{sine-sp}}
\label{transit-sp}
\end{figure}

\begin{figure}
 \vspace{0pt}
 \includegraphics[angle=-90,width=85mm]{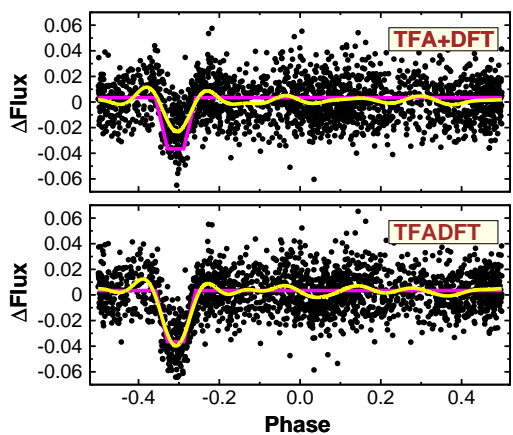}
 \caption{Folded light curves for the object with the frequency 
          spectra shown in Fig.~\ref{transit-sp}. The input trapezoidal 
	  transit signal is shown by pink line. The best $10$-th 
	  order Fourier fit is overpolotted by yellow line. The 
	  partial time series model (upper panel) yields shallower 
	  transit depth albeit both methods find the signal with 
	  the same significance.}
\label{transit-lc}
\end{figure}

The results of the same type of tests for the transit signal (test 
signal \#2 in Table~\ref{test-signals}) are shown in 
Figs.~\ref{transit-dr},\ref{transit-sp} and \ref{transit-lc}. For 
the detection statistics a similar pattern to that of the sine 
test signal is observable, although the difference between the two 
methods is considerably smaller. This effect is likely due to 
the increased flexibility originating from the $10$-th order Fourier 
fit used to model the transit signal. This leads to higher 
false signal pick-up rates and a decrease in the importance 
of the better systematics filtering of the partial model. 
We expect that if the signal were modeled instead with a boxcar, 
which uses fewer parameters and is thus less flexible, then {\sc tfa+dft} 
would provide an even greater enhancement over {\sc tfadft}, closer 
to what was seen for the sinusoidal case.  

The frequency spectra of a selected target (different from the one 
used in the sine test signal case) is shown in Fig.~\ref{transit-sp}. 
The difference between the two methods is insignificant. The folded LCs 
(Fig.~\ref{transit-lc}) are also similar but the decrease in the 
signal amplitude (i.e., transit depth) for the partial model is rather 
significant, similarly to the sinusoidal test signal case.

%

%

Not only that the partial model detects signals with higher SNR, it 
also has the ability to detect the signal exclusively (i.e., when the 
full model prefers another frequency, which corresponds either to some 
systematics or some real signal present in the data in which the test 
signal is injected). We examine the mutually exclusive detections 
in the two types of models. Since these detections come into play 
usually at low SNR values, we pose only the frequency condition as the 
sole criterion for detection. The result for the two types of signal 
and different TFA template numbers are shown in Table~\ref{exclusive-detect}. 

We see that the partial model leads to significantly larger number of 
exclusive detections. However, two caveats should be mentioned here. 
First, most of the `detections' have rather low SNR, leading to no 
detections in real applications. Roughly only one third of the exclusive 
detections for the partial models have SNR$>8$. For the full model 
the situation is worse and most of the exclusive detections have 
alias counterpart (with higher SNRs) in the partial model tests. 
Second, there might be also long-period variations in the stars, some 
of these variations could be intrinsic or long-term systematics. These 
might be preferred by the full model, leading to no detection of the 
injected signal. Whether or not these signals preferred by the full 
model are real, can be decided only by careful case-by-case studies, 
e.g., by running the analysis on varying TFA template numbers 
(Kov\'acs \& Bakos~\cite{kovacs2007}), inspecting the light curves and 
checking other stars with similar periods. Our variable star works on 
various datasets show that the loss rate of long-period variables is 
small, and can be handled in the way mentioned (e.g., D\'ek\'any \& 
Kov\'acs~\cite{dekany2009}, 
Szul\'agyi, Kov\'acs, \& Welch~\cite{szulagyi2009}, 
Kov\'acs et al.~{2014}).

%
\begin{table}
 \centering
 \begin{minipage}{140mm}
  \caption{Mutually exclusive detections}
  \label{exclusive-detect}
  \scalebox{1.00}{
  \begin{tabular}{cccc}
  \hline
   Signal  & $M$ & $R_{\rm part}$ & $R_{\rm full}$ \\
 \hline
 $\#1$ & 200 & 0.111 & 0.020 \\
       & 400 & 0.121 & 0.017 \\
 $\#2$ & 200 & 0.131 & 0.020 \\
       & 400 & 0.077 & 0.024 \\ 
\hline
\end{tabular}}
\end{minipage}
\begin{flushleft}
{{\bf Notes:} $R_{\rm part}$ denotes the ratio of the number of exclusive 
detections in the partial ({\sc tfa+dft}) model to the total number of 
objects tested (i.e., to $300$). Similarly, $R_{\rm full}$ refers to the 
same type of detections for {\sc tfadft}. No SNR condition is posed.}  
\end{flushleft}
\end{table}

%
\subsection{Ensemble test: results based on K2}
Most of the current works advocating simultaneous systematics 
filtering and signal search focus their attention on the data gathered 
by the K2 mission. Therefore (as suggested by the referee), it is of 
considerable importance to investigate if the overall better/similar 
performance of separate systematics filtering ({\sc tfa+dft}) as indicated 
by the HATNet data survives also for the time series of the K2 mission. 
The selection procedure for the K2 test dataset has been described in 
Sect. 3.2.

Sinusoidal signal tests show quite clearly that unlike for the HATNet data, 
the detection rate is not a monotonic function of the TFA template number 
($N_{\rm TFA}$) for the K2 data. Therefore, here we plot the detection rate 
as a function of $N_{\rm TFA}$ at a fixed lower bound of $SNR_{\rm min}=6$ 
for the SNR of the frequency spectra. (The relative topology of the detection 
rates is not affected in an essential way by changing $SNR_{\rm min}$ in both 
directions.) 

\begin{figure}
 \vspace{0pt}
 \includegraphics[angle=-90,width=85mm]{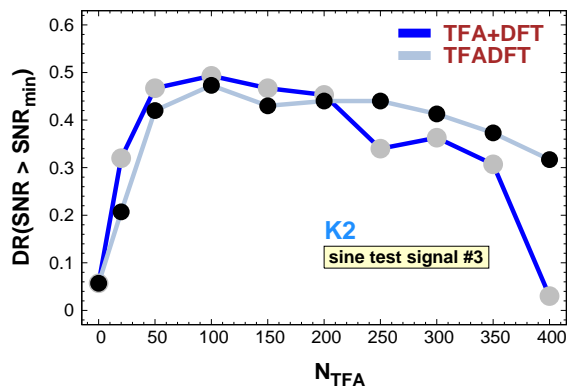}
 \caption{Detection ratio as a function of the number of the TFA 
          templates for the injected sinusoidal test signal \#3 
          (Table~\ref{test-signals}) in the subset of the K2/Campaign 1 
	  data (see Sect.~3.2). Simultaneous systematics filtering and 
	  signal search ({\sc tfadft}) performs somewhat poorer, 
	  except for higher template numbers, where the detection ratio 
	  saturates/decreases.}
\label{ntfa-det-dft}
\end{figure}

The result for the sinusoidal test signal (\#3 of Table~\ref{test-signals}) 
is shown in Fig.~\ref{ntfa-det-dft}. The decrease of the detection rate (DR) 
both for the full-fledged and for the partial search is well exhibited. The 
full-fledged search have better statistics at the high $N_{\rm TFA}$ end but 
the downward trend for this type of search is also obvious. Before and 
shortly after the optimum detection rate, {\sc tfa+dft} and {\sc tfadft} 
behave similarly, with a slight preference toward higher detection rates for 
{\sc tfa+dft}. This behavior is similar to what we found for the HATNet data. 

%

%
\begin{figure}
 \vspace{0pt}
 \includegraphics[angle=-90,width=85mm]{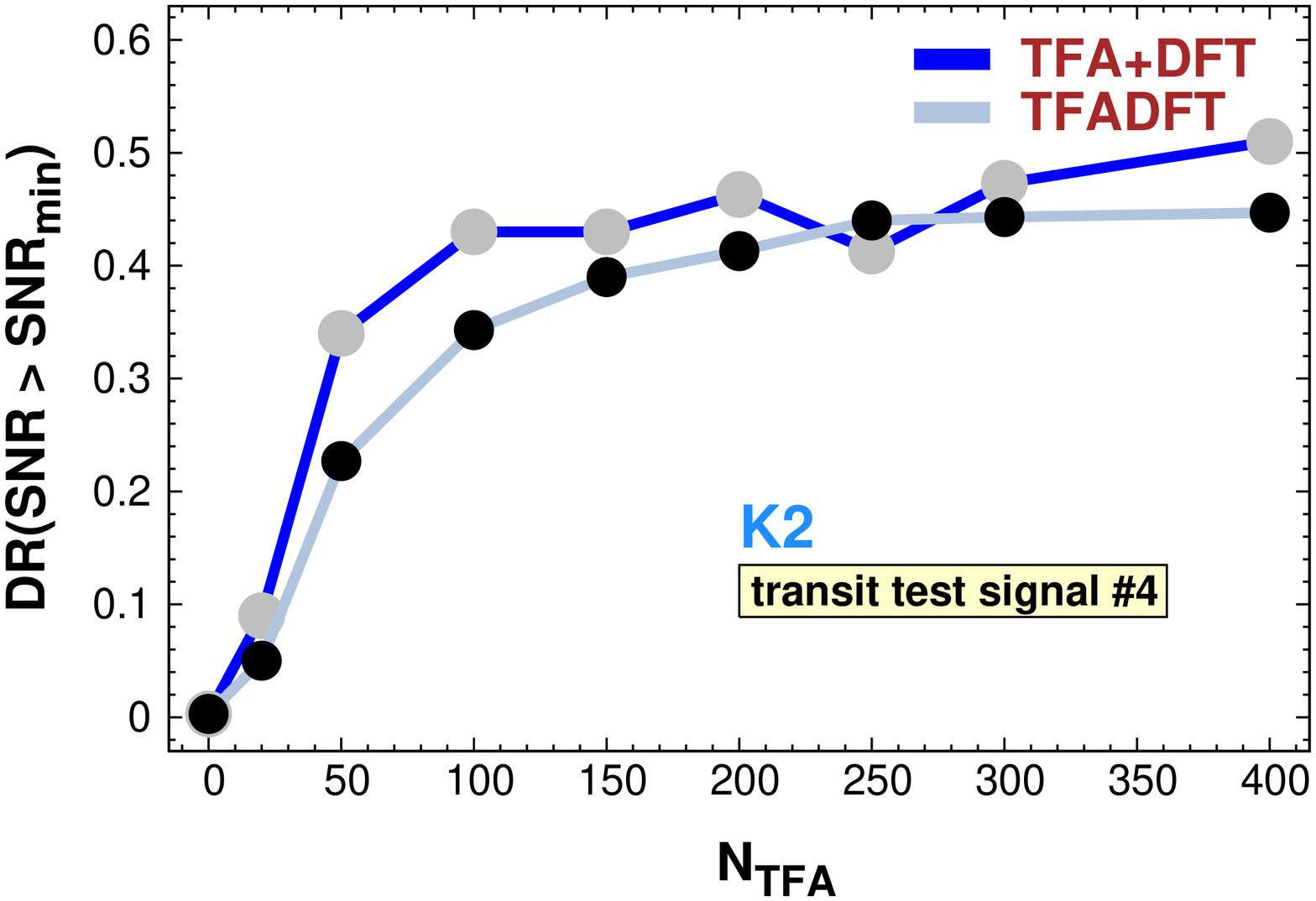}
 \caption{Detection ratio as a function of the number of the TFA 
          templates for the injected transit test signal \#4 
          (Table~\ref{test-signals}) in the subset of the K2/Campaign 1 
	  data (see Sect.~3.2). Simultaneous systematics filtering and 
	  signal search ({\sc tfadft}) performs poorer, except 
	  for higher template numbers, where the detection ratio saturates.}
\label{ntfa-det-bls}
\end{figure}

%
\begin{figure}
 \vspace{0pt}
 \includegraphics[angle=-90,width=85mm]{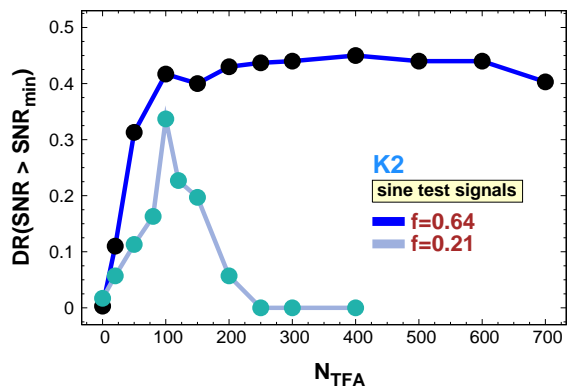}
 \caption{Testing the dependence of the detection ratio for injected 
          sinusoidal signals in the K2 data. Test signals \#5 and \#6 of 
	  Table~\ref{test-signals} are used in the `systematics filtering 
	  first, frequency search thereafter' ({\sc tfa+dft}) mode. The long 
	  period signal has a brief template range around $N_{\rm TFA}\sim 100$ 
	  where the detection ratio is optimal. The detection rate for the 
	  short period signal also saturates around this value but unlike 
	  the long period signal it remains around the optimum rate also for 
	  rather high TFA template numbers.} 
\label{ntfa-max-dft}
\end{figure}

%
\begin{figure}
 \vspace{0pt}
 \includegraphics[angle=-90,width=85mm]{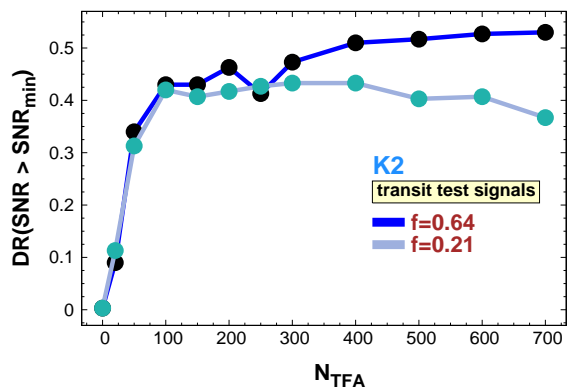}
 \caption{As in Fig.~\ref{ntfa-max-dft} but for the transit signals 
 \#4 and \#7 of Table~\ref{test-signals}. Unlike the sinusoidal signals, 
 transit signals remain optimally detected also for high TFA template numbers.}
\label{ntfa-max-bls}
\end{figure}

The detection rate for the transit signal (\#4 of Table~\ref{test-signals}) 
shows a similar behavior in the low $N_{\rm TFA}$ regime 
(see Fig.~\ref{ntfa-det-bls}). However, in the high $N_{\rm TFA}$ regime 
it behaves quite differently from the sinusoidal signal. The detection rate 
slightly increases for both search methods, a behavior observed both for 
the sinusoidal and for the transit test signals for the HATNet data. We 
conclude from these tests that, for the K2 data and for transit type signals, 
the partial model works similarly (or better) than the full-fledged model 
for a large range of TFA template numbers. For sinusoidal signals the same 
is true, except in the higher template regime, where the full-fledged method 
tends to outperform the partial model (albeit at a sub-optimal detection 
ratio). 

Although the discussion and deeper testing of the non-monotonic nature of 
the detection rates for sinusoidal signals does not belong to the focus of 
this work, to have a somewhat better glimpse on the problem, we perform 
additional tests including the sinusoidal signals \#5, \#6 and transit 
signals \#4, \#7 of Table~\ref{test-signals}. (We note that these signals 
are constructed for testing the effect of the period, therefore, the same 
type of signals have the same amplitude/transit depth.) Since we aimed at 
testing CPU-demanding high template numbers, and the full-fledged search 
tend to yield similar results to the search based on the partial model, 
we restrict ourselves to the latter. Figs.~\ref{ntfa-max-dft},~\ref{ntfa-max-bls} 
confirm our earlier conclusion on the sensitivity of the detection rates 
of the sinusoidal signals on the template number and the insensitivity 
of the transit signals on the this parameter. However, it is important 
to note that for shorter signal periods the detection ratio for sinusoidal 
signals is significantly less sensitive to the number of TFA templates. 
Interestingly, the detection rates for transit type signals show fairly 
good stability even at higher template rates, although shorter period 
signals perform better (i.e., their detection rates tend to increase with 
the template number). Both for the sinusoidal and for the transit signal 
template numbers of $100$--$150$ yield close to optimum (maximal SNR) 
detection ratios. It is interesting to note that 
Foreman-Mackey et al.~(\cite{foreman2015}) use similar number of 
eigen LCs from their PCA set derived on the Campaign~1 data.     
 
Concerning the underlying cause of the depression of the detection rate 
for longer period sinusoidal signals, it is strongly suspected that the 
effect is attributed to the lower noise of the K2 data that enables the 
showing up of many physical variables among the template members (and 
obviously, the probability of picking up variables in the template set 
increases with the increase of the template numbers). The most likely 
impostors are the spotted variables, since they have close to sinusoidal 
light curves, cover a wide range of periods and some level of stellar 
activity is a generic property of main sequence stars (an observation, 
strongly supported by the recent discovery of a large number of 
rotational/spotted variables from the Kepler database -- see McQuillan, 
Mazeh \& Aigrain~\cite{mcquillan2014}). For the HATNet data this is 
not a significant issue, since the noise floor for HATNet is higher, 
which blurs the physical signal for most of the low amplitude spotted 
stars. The fact that the transit signals have a much better survival 
rates also implies that most of the `signal killers' are sinusoidal 
variables.

%
\section{Conclusions}
Identifying systematic effects in astronomical time series and correcting 
them without jeopardizing (but rather improving) the detection power of 
various signal search algorithms is a prime topic in contemporary efforts 
to find small transiting extrasolar planets and investigate low-amplitude 
stellar variability. Without knowing the exact time dependence of the 
systematics, we need to find the best method to disentangle the signal 
component from these, physically uninteresting effects. One obvious choice 
to retain both the signal content and, at the same time, filter out systematics, 
is to conduct a parameter search that includes both effects simultaneously. 
Although it is clear that at the end of any signal search one has to resort 
to a full model fit, it is unclear whether carrying out a full fit for 
systematics while searching for signals will not yield false detections due 
to the increased freedom in the fit. In particular, in the course of the 
frequency search, one has to try usually a huge number of cases. There might 
exist frequency bands, where a considerable part of the systematics is well 
approximated by the Fourier series or by some matched filter we use for the 
representation of the signal being searched for. In these cases the statistics 
used to construct the frequency spectrum might falsely indicate that there 
is a signal in that frequency band. Above all these, of course, there is 
also the computational deterrence in performing such a multiparametric 
search, primarily because of the strongly variable nature of the goodness 
of fit statistics on the test frequency.        

In this paper we examined the performance of the frequency search methods 
based on full (systematics$+$signal) model fits. Four recent papers 
(Aigrain et al.~\cite{aigrain2015}; Foreman-Mackey et al.~\cite{foreman2015}; 
Angus et al.~\cite{angus2015} and Wang et al.~\cite{wang2015}) advocate  
this approach for the analysis of the photometric time series obtained by 
the K2 mission (the two reaction wheels program of the {\it Kepler} 
satellite). Because no counter tests with separately applied systematics 
filtering and signal search have been presented in those papers, we think 
it is important to execute such a test before we start a broader application 
of the full-fledged model method. 

The tests presented in this paper (based on purely artificial data and 
signals injected in a sample of observed light curves from the HATNet 
project and from the Campaign 1 data of the K2 mission) clearly 
show that for signal search the partial model (systematics fitting only), 
is preferable over the full model search. This statement is based 
on the signal-to-noise ratio of the frequency spectra and tests performed 
on sinusoidal and transit-like periodic signals. For the HATNet data, the 
advantage of the partial model fit is especially visible for the sinusoidal 
test signals, where the detection rate is more than $10$\% higher for the 
partial model. For transit-like signals this difference goes down to a few 
percent. 

Because of the considerably lower noise level of the K2 data, the above 
conclusion should be expanded somewhat for signal shapes similar to those 
produced by intrinsic stellar variability (primarily by spot modulation due 
to stellar rotation). Depending on the period of the target, the detection 
ratio might decrease considerably for longer period sinusoidal signals for 
systematics filtering co-trending time series (template light curves) greater 
than $\sim 100$--$150$. With large template numbers the chance of finding 
intrinsic variables among them with periods close to that of the target, 
increases. This may lead to filtering out also the signal, not only the 
systematics. Transit signals are more robust in this respect, since similar 
quasi-coincidence with the co-trending template sample is far less likely. 
Although full-fledged search may prolong the survival rate somewhat in the 
regime of high template numbers, it does not lead to the detection of 
additional signals. The maximum detection rates resulting from the signal 
search based on partial time series modeling were never exceeded by the 
full-fledged models on any of the datasets tested.    

Considering the possibility that a proper choice of basis (e.g., 
TFA template) functions and weighting of the signal and systematics 
parts of the model -- based, e.g., on some more fundamental principles, 
such as the Bayesian inference -- may still improve the situation for 
the full-fledged search, we add the following. 

It is possible that a full-fledged method which includes a penalty 
for over-fitting the data, in particular which penalizes fitting 
periodic signals at frequencies where the systematics are found to 
have significant power and where the data would be equally well modeled 
using only the systematics filter, may have better performance at low 
S/N than the full-fledged method tested here.  The filtering techniques 
presented, for example, by Roberts et al.~(\cite{roberts2013}) or by 
Smith et al.~(\cite{smith2012}) might be amenable to such an extension. 
Although we cannot predict the effect of the above extension of the 
full-fledged model on the effectiveness of the frequency search, based 
on the tests related to the subject of the paper we doubt that such a 
method would perform substantially better at low S/N than the two-step 
procedure. 

Finally, the full model search is (obviously) always more time consuming, 
the one presented in this paper often by a factor of ten, depending 
slightly on the method of implementation. However, it is clear that 
in any period search influenced by systematics, the final step must be a 
full model fit (which is inexpensive). Once the signal is found, a 
full model fit will handle all constituents of the input time series 
properly and lead to the reconstruction of the signal by alleviating the 
effect of `signal squeezing' caused by the partial model fit.

\begin{acknowledgements}
We would like to thank to Daniel Foreman-Mackey for the enlightening 
correspondence in the early phase of this project. This work has been 
started during G.~K.'s stay at the Physics and Astrophysics Department 
of the University of North Dakota. He is indebted to the faculty and 
staff for the hospitality and the cordial, inspiring atmosphere. We 
appreciate the critical but constructive comments of the anonymous 
referee. This paper makes use of data from the HATNet survey, the 
operation of which has been supported through NASA grants NNG04GN74G 
and NNX13AJ15G. J.~H.~acknowledges support from NASA grant NNX14AE87G. 
\end{acknowledgements}

\bibliographystyle{aa} 

\end{document}